\documentclass[prb,preprint,showpacs,preprintnumbers,amsmath,amssymb]{revtex4}

\usepackage{graphicx}
\usepackage{dcolumn}
\usepackage{bm}

\begin{document}

\preprint{Submitted to PRB}

\title{Wavelength Tunability of Ion-bombardment Induced Ripples on Sapphire}

\author{Hua Zhou}
\author{Yiping Wang}
\author{Lan Zhou}
\author{Randall L. Headrick}\email{rheadrick@uvm.edu}
\affiliation{Department of Physics, University of Vermont, Burlington, VT 05405 USA}
\author{Ahmet S. \"{O}zcan}
\author{Yiyi Wang}
\author{G\"{o}zde \"{O}zaydin}
\author{Karl F. Ludwig Jr.}
\affiliation{Department of Physics, Boston University, MA 02215 USA}
\author{D. Peter Siddons} \affiliation{National Synchrotron Light Source, Brookhaven
National Laboratory, Upton, NY 11973 USA}

\date{\today}

\begin{abstract}
A study of ripple formation on sapphire surfaces by 300 - 2000 eV
Ar$^+$ ion bombardment is presented. Surface characterization by
\textit{in-situ} synchrotron grazing incidence small angle x-ray
scattering and \textit{ex-situ} atomic force microscopy is performed
in order to study the wavelength of ripples formed on sapphire
(0001) surfaces. We find that the wavelength can be varied over a
remarkably wide range -- nearly two orders of magnitude -- by
changing the ion incidence angle. Within the linear 
theory regime, the ion induced viscous flow smoothing mechanism
explains the general trends of the ripple wavelength at low
temperature and incidence angles larger than 30$^\circ$. In this
model, relaxation is confined to a few-nm thick damaged surface
layer. The behavior at high temperature suggests relaxation by
surface diffusion. However, strong smoothing  is inferred from the
observed ripple wavelength near normal incidence, which is not
consistent with either surface diffusion or viscous flow relaxation.
\end{abstract}

\maketitle

\section{Introduction}
Energetic particle bombardment on surfaces is known to produce one-dimensional
(ripples or wires) and zero-dimensional (dot) structures at the submicron or
nano-scale by a self-organization process. Recently, significant
experimental and theoretical effort has been expended in order to
develop ion bombardment patterning methods for the production of
periodic nanostructures on various substrates.~\cite{Valbusa3,
Erlebacher4, Chason5, Carter29, Datta6, Rusponi7, Cuerno15,
Makeev16} These studies have demonstrated the potential to tailor
surface morphology and related surface properties for novel
optoelectronic and spintronic applications.~\cite{Facsko8, Moroni40}
In addition, recent work has provided new insight into the
mechanisms of the instability-driven self-organization
process.~\cite{Makeev2}

A significant milestone in our understanding of the origins of a
self-organized ripple topography formed by ion sputtering is the
work of Bradley and Harper (BH) in which they proposed a linear
continuum equation to describe the main features of ripple
formation.~\cite{BH10} The main idea of BH is that smoothing and
roughening processes have different wavelength dependence, leading
to a preferred wavelength where the surface amplitude grows the most
rapidly.

However, certain experimentally observed features, such as the
saturation of the ripple amplitude and the appearance of kinetic
roughening are not predicted by the linear BH
theory.~\cite{Erlebacher12, Eklund52} An extension of the linear BH
theory into the non-linear regime has been proposed in order to
avoid these shortcomings,~\cite{Makeev2} resulting in  a noisy
version of the Kuramoto-Sivashinsky equation:
\\
\begin{eqnarray}\label{KS-1}
  \frac{\partial{h}}{\partial{t}}=-\upsilon_{0}+\zeta\partial_{x}{h}+\xi_{x}\partial_{x}{h}\partial_{x}^{2}{h}+\xi_{y}\partial_{y}{h}\partial_{y}^{2}{h}+\nu_{x}\partial_{x}^{2}{h}+\nu_{y}\partial_{y}^{2}{h}-K_{xx}\partial_{x}^{4}{h}\nonumber
  \\-K_{yy}\partial_{y}^{4}{h}-K_{xy}\partial_{x}^{2}\partial_{y}^{2}{h}+\frac{\lambda_{x}}{2}(\partial_{x}{h})^2+\frac{\lambda_{y}}{2}(\partial_{y}{h})^2+\eta{(x,y,t)}~.
\end{eqnarray}
\\
The  $\upsilon_{0}$ term, which represents the average erosion
rate of the unperturbed planar surface can be neglected in Eq. (1),
since it does not affect the process of ripple formation. The
surface height $h$ is then in a coordinate system that moves with
the average surface during the erosion process. $\eta{(x,y,t)}$ is a
Poisson noise term related to random fluctuations, uncorrelated in
space and time, in the flux of the incoming ions.

Within this theory, ion sputtering produces periodic modulated
features (correlated lateral ordering) that arises from a
competition of a roughening instability mechanism and surface
relaxation. A roughening mechanism that often dominates the surface
morphology is curvature-dependent sputtering, which is based on the
linear cascade approximation first proposed by
Sigmund.~\cite{Sigmund11} However, certain compounds, such as GaSb
and InP,~\cite{Facsko8,Frost41} and elemental materials with
refractory-metal seeding,~\cite{Ozaydin32} may also exhibit island
agglomeration of excess elements by a process related to
preferential sputtering.

In contrast, a wider range of relaxation mechanisms have been
proposed in order to explain various experimental observations: (i)
Surface diffusion (SD) mediated smoothing has been proposed to
explain the temperature and ion flux dependence of ripple wavelength
in the high temperature regime.~\cite{Erlebacher4,MacLaren13} (ii)
The surface erosion smoothing (SES) model in which higher-order terms of the erosion process
produce a smoothing which mimics  surface diffusion.~\cite{Makeev16} (iii) The ion-enhanced viscous flow
relaxation (IVF) model, which considers surface-confined viscous flow
driven by surface tension as the dominant smoothing mechanism for
any material with a disordered surface layer with reduced
viscosity.~\cite{Umbach14} Mayr et al. have emphasized the role of
flow of point defects and of thermal-spike induced local melting in
the mechanisms of radiation-induced viscous flow in the 0.1-1 KeV
range.~\cite{Mayr34} However, determining which relaxation mechanism
(SD, SES, IVF, etc.) is dominant for various amorphous or
crystalline substrates, surface temperatures, and ion beam
parameters, still remains the subject of vigorous research.

\section{Linear Theories}
In the theoretical approach of Makeev and
Barab\'{a}si,~\cite{Makeev2} a general continuum equation which
describes the evolution of surface morphology during ion sputtering
is proposed. Eq. 1 incorporates the
major features of ion-bombardment induced ripple formation and
kinetic roughening. In the early stage of ion sputtering, Eq. (1)
reduces to a BH-type linear theory when
$\lambda_{x}$=$\lambda_{y}$=0, $\xi_{x}=\xi_{y}$=0. In this article,
our discussions of the wavelength tunability of ion-bombardment
induced sapphire ripples are emphasized within the linear regime.
The linear terms with coefficients $\nu_{x}$ and $\nu_{y}$ represent
the curvature dependent ion erosion rates, and K$_{xx}$ and
K$_{yy}$ are coefficients representing the surface smoothing terms.

Linear stability analysis indicates that the establishment of a
periodic ripple structure across the surface depends on the balance
between the curvature dependent roughening and surface smoothing
mechanisms.~\cite{BH10} Two modes of rippled morphology can be
induced by ion bombardment, with ripple wavevectors parallel or perpendicular to the
projection of the ion beams. Regardless of the respective smoothing
mechanisms (SD, SES, IVF), the wavelength of ion sputtered ripples
with orthogonal orientations $\ell_{x}$ (parallel) and $\ell_{y}$ (perpendicular) are generally
expressed as:
\\
\begin{eqnarray}\label{ripple wavelength}
   \ell_{x}=2\pi\sqrt{\frac{2K_{xx}}{|\nu_{x}|}}~,
   \\\ell_{y}=2\pi\sqrt{\frac{2K_{yy}}{|\nu_{y}|}}~.
\end{eqnarray}
\\
The minimum between $\ell_{x}$ and $\ell_{y}$ determines which orientation dominates the
 ion-induced ripple topography. In Eq. (2) and (3), the
coefficients $\nu_{x}$, $\nu_{y}$ for curvature dependent roughening
terms are given (following Ref. 11) by :
\\
\begin{equation}
\nu_{x}=Fd\frac{d_{\sigma}^{2}}{2f^{3}}\left\{2d_{\sigma}^{4}s^{4}-d_{\sigma}^{4}d_{\mu}^{2}s^{2}c^{2}+d_{\sigma}^{2}d_{\mu}^{2}s^{2}c^{2}-d_{\mu}^{4}c^{4}\right\},
\end{equation}
\begin{equation}
\nu_{y}=-Fd\frac{c^2d_{\sigma}^2}{2f}~.
\end{equation}
\\
In the expressions above, the terms are defined as:
\\
\begin{eqnarray}
      d_{\sigma}\equiv\frac{d}{\sigma}, d_{\mu}\equiv\frac{d}{\mu},\nonumber
    \\s\equiv{\sin(\theta)}, c\equiv{\cos(\theta)},\nonumber
    \\f\equiv{d_{\sigma}^{2}s^2+d_{\mu}^{2}c^2},\nonumber
    \\F\equiv\frac{J\bm{\varepsilon}pd}{\sigma\mu\sqrt{2\pi{f}}}e^{-d_{\sigma}^{2}d_{\mu}^{2}c^2/2f},
\end{eqnarray}
\\
where F is a coefficient relating to the local sputter yield $Y(\theta$)
as:
\\
\begin{equation}
F=\frac{JY(\theta)}{nc}~.
\end{equation}
\\
In addition, $d$ is the ion energy deposition depth, $\sigma$ and
$\mu$ are ion energy distribution widths parallel and perpendicular
to the incoming ion beams, J is the ion flux per area, $p$ is a material constant
depending on the surface binding energy U$_{0}$ and scattering
cross-section,~\cite{Sigmund11} and $n$ is the atomic density of the
substrate.

For the SD smoothing mechanism, thermally activated surface
diffusion induces surface smoothing during ion sputtering. If
surface self-diffusion is isotropic, then
\\
\begin{equation}
    K_{xx}=K_{yy}=K_{SD}=\frac{D_{s}\gamma\rho}{n^2k_{B}T}~.
\end{equation}
\\
Here, D$_{s}$ is the surface self-diffusivity, which has an Arrhenius
temperature dependence D$_{s}$=D$_{0}exp(-\Delta{E}/k_{B}T)$. The
surface tension (surface free energy per unit area) is $\gamma$,
and $\rho$ is the areal density of diffusing atoms.~\cite{BH10}

For the SES model, the erosion smoothing process is assumed to
dominate over other smoothing mechanisms. Since it is not
temperature dependent, it is thought to dominate at low temperatures
in some cases. Here, K$_{xx}$=K$_{xx, SES}$ and K$_{yy}$=K$_{yy,
SES}$, which are anisotropic with respect to the direction of the
incoming ion beams. The coefficients representing surface erosion
smoothing K$_{xx,SES}$ and K$_{yy,SES}$ are given (in Ref. 11) by :
\\
\begin{eqnarray}
  K_{xx,SES}=F\frac{d^{3}}{24}\frac{1}{f^{5}}\left\{-4(3d_{\sigma}^{2}s^2f+d_{\sigma}^{6}s^4)f^2+d_{\sigma}^{2}c^2(3f^2+6d_{\sigma}^{4}s^2f+d_{\sigma}^{8}s^4)f\right\}\nonumber
  \\+F\frac{d^{3}}{24}\frac{1}{f^{5}}\left\{2(d_{\mu}^{2}-d_{\sigma}^{2})c^2(15d_{\sigma}^{2}s^2f^2+10d_{\sigma}^{6}s^4f+d_{\sigma}^{10}s^6)\right\},
\end{eqnarray}

\begin{equation}
  K_{yy,SES}=F\frac{d^{3}}{24}\frac{c^2}{f}\frac{3d_{\sigma}^2}{d_{\mu}^2}~.
\end{equation}
\\

For the IVF model, the ion-enhanced surface viscous flow within a
thin ion-damaged layer dominates the surface smoothing. Hence,
\
\begin{equation}
    K_{xx}=K_{yy}=K_{IVF}=\frac{\gamma{d^3}}{\eta_{s}}~.
\end{equation}
\\
Here, the ion-enhanced surface viscosity $\eta_{s}$ and the surface
tension are assumed to be constant and isotropic. The depth of the
damaged layer is taken to be equal to the ion penetration depth
$d$.~\cite{Umbach14}

In the following sections, the data analysis and discussion of the
wavelength tunability of ion-bombardment induced sapphire ripples
are based on Eqs. 2-11 for the ripple wavelength.  Additional
smoothing mechanisms beyond SD, SES and IVF are considered in
section V.

\section{Experimental}

The ripples are produced on sapphire (0001) in a custom surface
x-ray ultra-high vacuum chamber ($10^{-9}$ torr base pressure)
installed at x-ray beamline X21A1 at the National Synchrotron Light
Source (NSLS) at Brookhaven National Laboratory. More details about
this real-time surface x-ray characterization facility for dynamic
processing are described elsewhere.~\cite{Ozcan42} Ion beam
sputtering is performed by either a 3-grid RF plasma ion source or a
Phi model 04-192 sputter ion gun. The RF plasma ion source is
operated at ion energies ranging from 300 eV to 1000 eV at a flux of
2$\times10^{14}$ ions/cm$^{2}$/sec with a background Ar$^{+}$
pressure 4$\times10^{-4}$ torr. The irradiation size
of the ion beam with a uniform flux (3 cm) is large enough to cover the
entire sample surface.   The Phi sputter ion gun is operated at ion
energies from 500 eV to 2000 eV at a flux of 
1$\times10^{13}$ ions/cm$^{2}$/sec with a background Ar$^{+}$
pressure 1$\times10^{-4}$ torr. The sample surface
temperature is adjusted from 300 K to 1050 K, and monitored by an
infrared pyrometer. The chamber is also equipped with reflection high energy electron diffraction (RHEED), which is used to determine surface crystallinity.  A contact mode Digital Instruments Nanoscope-E AFM is used for \textit{ex-situ} surface morphology imaging.

The  x-ray flux after the Si (111) monochromator crystal is
2$\times10^{12}$ photons/sec at a wavelength of $\lambda$=1.192
{\AA} with a beam-size of 0.5 mm $\times$ 0.5 mm. In the schematic
representation of the x-ray measurement geometry shown in Fig. 1,
the z axis is always taken to be normal to the sample surface, and
the y axis along the projection of the incident x-ray beams onto the
surface. $\textbf{k}_{\textbf{i}}$ and
$\textbf{k}_{\textbf{f}}$ are the wave vectors of the incoming and
scattered x rays, respectively. The components of the scattering
momentum transfer
$\textbf{Q}=\textbf{k}_{\textbf{f}}-\textbf{k}_{\textbf{i}}$ can be
expressed by the glancing angles of incidence ($\alpha_{i}$) and
exit ($\alpha_{f}$) with respect to the surface (x-y plane), and the
in-plane angle $\psi$.

In the GISAXS geometry, the incident or exit x-ray beams are fixed
near the critical angle for total external reflection ($0.2^{\circ}$
for sapphire). A 320-pixel linear position sensitive detector (PSD)
is positioned along x axis at the angle  $\alpha_{f}$ with respect to the
surface, in order to collect in-plane scattered x rays. In terms of scattering
momentum transfer, the PSD acquires a range of $Q_{x}$ at a constant
$Q_{z}$ and $Q_{y}$ (typically, $Q_{z}$=0.92 $nm^{-1}$,
$\bigtriangleup$$Q_{x}$$\approx$1 nm$^{-1}$ and
$Q_{y}$=3$\times$10$^{-3}$ nm$^{-1}$). Time-resolved GISAXS provides access to the evolving wavelength,
shape and amplitude of surface ripples. For a full reciprocal
mapping ($Q_{x}$ vs. $Q_{z}$), an $\alpha_{i} = \alpha_{f}$ reflection
mode scan is performed.

\section{Results}

In this part, subsections A through D describe a series of systematic
investigations of the dependance of sapphire ripple characteristics (wavelength, orientation), on experimental parameters, including ion
energy, ion incidence angle and temperature.

\subsection{Ripple Formation by Ion Sputtering}
Fig. 2 displays a Q$_{x}$ vs. Q$_{z}$  GISAXS scan of the rippled
sapphire surface after $45^{\circ}$ off-normal 500 eV $Ar^{+}$
bombardment. First and second order satellite streaks are readily observed, indicating a periodic surface morphology along the x-direction. In this image, the specular reflected x-rays are cut off by
a narrow beam stop attached to the PSD.  This provides  better sensitivity to
off-specular diffuse scattering pattern.

Fig. 3 shows an AFM image of the as-irradiated sapphire surface. A
ripple morphology is clearly visible. A wavelength of 32 nm is
obtained from the image, which corresponds well with that acquired
from GISAXS in Fig. 2.

Fig. 4 shows real-time GISAXS intensities plotted versus scattering
momentum transfer $Q_{x}$. The scans are equivalent to a linear slice of the image in Fig. 2
at a constant vertical component of the scattering momentum transfer
Q$_{z}$=0.92 nm$^{-1}$, except that that an offset condition ($\alpha_{i} \ne \alpha_{f}$) was used so that $Q_{y}$=3$\times$10$^{-3}$ nm$^{-1}$. This was done in order to eliminate the need for a specular beam stop.  Scans are shown at 10-minute intervals
during $45^{\circ}$ off-normal 500 eV $Ar^{+}$ bombardment.  At time
t=0, the initial roughness of the sapphire surface only produces a
single peak in the diffuse scattering (circles). Two satellite peaks,
resulting from lateral correlated roughness become visible after 10
minutes. The ripple wavelength, $\ell_{x}=2\pi/\Delta{Q_{x}}$,
remains almost constant during ion irradiation, but the peak
intensity continues to increase as a result of an increase in ripple
amplitude.

It is also clear in Fig. 4 that the two satellite peaks develop in
an unequal way as irradiation proceeds. After 30 minutes, the peak
on the positive $Q_{x}$ side is noticeably larger than the one on the
negative $Q_{x}$ side. At 40 minutes, the larger peak is several
times more intense than the smaller one. This diffuse intensity
asymmetry was also observed in the GISAXS study of ion-eroded
SiO$_{2}$ by Umbach et al.~\cite{Umbach14}

Fig. 5 shows an asymmetric saw-tooth model profile, which is used as
a simplified approximation to the ripple shape. Here, the parallel
component of the incident ion beam is along the -x direction, as
defined above. We note that the term $\partial{h}/\partial{x}$ in
Eq. (1), representing the surface local slope, has opposite signs at
different sides of the solid saw-tooth. Thus, an off-normal
incidence will produce different erosion rates on
positive and negative slopes. The model predicts that
the unbalanced erosion will make ripples move like waves across the
surface in a direction opposite to  the projection of the incident beam along the surface.~\cite{BH10} However, we note that a recent study of ripples formed on ion-bombarded glass surfaces showed forward propagation of ripples.~\cite{Alkemade55} Nonlinear terms of the form
$(\partial{h}/\partial{x})(\partial^{2}{h}/\partial{x}^{2})$ produce
the asymmetric shape, which is  observed in our x-ray diffuse scattering measurements.~\cite{Makeev2}
Therefore, the appearance of an asymmetric GISAXS pattern  indicates
the onset of this lowest order non-linear term.

\subsection{Ripple Wavelength Variation with Ion
Energy}

Fig. 6 shows the observed dependence of ripple wavelength $\ell_{x}$
on ion energy $\bm{\varepsilon}$ for ion sputtered sapphire at low
temperature 300 K (a) and high temperature 1000 K (b),
respectively. This series of sapphire ripples are obtained at
45$^{\circ}$ off-normal ion incidence. In Fig. 6(a), square and
circle symbols represent the wavelength of ripples produced by the
high-flux RF plasma ion source and the low-flux ion gun,
respectively. The sapphire ripple wavelength increases with ion
energy at low temperature, which is consistent with observations for
ion eroded SiO$_{2}$, GaAs and Si surfaces.~\cite{Karen22, Vajo23,
Umbach14} The data obtained from high/low flux ion sources overlap
within experimental error at both 500 eV and 1000 eV, which
indicates that the ripple wavelength is independent of incident ion
flux at low temperature. A non-linear least squares fit gives a
power law coefficient with p=0.71 for the dependence of the
wavelength on ion energy ($\ell_{x}\sim\bm{\varepsilon}^{p}$). Also
plotted are curves corresponding to  p=1 and
0.5 for comparison. In Fig. 6(b), the ripple wavelength decreases
with ion energy. A non-linear least squares fit gives a power law
coefficient with p=-0.44. Also plotted are curves corresponding to p=-0.25 and -0.75 for comparison.

A general formula for a low temperature ripple wavelength along the
dominant orientation x-axis can be expressed from Eq. (2), (4) and
(9), based on the SES model (section II). Taking $|\nu_{x}|\sim{Fd}$
and K$_{xx,SES}\sim{Fd^3}$, we can obtain the dependence of the
wavelength on ion energy as:
\\
\begin{equation}\label{wavelength ion energy dependence1}
  \ell_{x}=2\pi\sqrt{\frac{2K_{xx,SES}}{|\nu_{x}|}}\sim{d}~.
\end{equation}
\\
The dependence of $d$ on $\bm{\varepsilon}$ is quantitatively
accessible with the aid of the ion-collison simulator
SRIM.~\cite{SRIM} It indicates that $d$ varies as
$\bm{\varepsilon^p}$ with p=0.48 for $\alpha$-sapphire at the
incidence angle of 45$^{\circ}$. The p=0.48 obtained from Eq. (12)
matches the observed wavelength-ion energy relation in Fig. 6(a)
reasonably well. However, a
quantitative analysis (as detailed in section IV-C) by Eq. (12)
predicts values of the ripple wavelengths that are an order of magnitude smaller than our measured values of $\ell_{x}$, indicating that the SES, which
contains no adjustable parameters, does not account for the observed
ripple wavelength at low
temperature.~\cite{Umbach14}

A specific expression for the ripple wavelength based on the
ion-enhanced surface viscous flow (IVF) model,~\cite{Orchard24} can
be derived from Eq. (2), (4), (7) and (11) as an extension to Eq.
(12). Inserting K$_{IVF}=\gamma{d^3}/\eta_{s}$, $|\nu_{x}|\sim{Fd}$ and
F=JY($\theta$)/nc into Eq. (2), we can get the ion energy dependence
for the IVF model as:
\\
\begin{equation}\label{wavelength ion energy dependence2}
\ell_{x}\sim2\pi{d}\sqrt{\frac{2\gamma{n}c}{JY(\theta)\eta_{s}}}~,
\end{equation}
\\
where the coefficient K$_{xx}$ of the IVF smoothing term replaces
K$_{xx}$ in Eq. (2). Simulations from SRIM at 
45$^{\circ}$ incidence indicate that Y($\theta$) varies as
$\bm{\varepsilon}^{\delta}$ with $\delta$=0.62. It is assumed from
previous work on low energy ion irradiation~\cite{Mayer25} that the
dependence of ion-enhanced surface viscosity on ion energy takes the
form J$\eta_{s}\sim\bm{\varepsilon}^{-\alpha}$ with $\alpha\approx$
1. Hence, $\ell_{x}\sim\bm{\varepsilon}^{p}$ where p=0.67 for
sapphire ripples. Here, the sapphire atomic density $n$ is
1.17$\times10^{29}$ atoms/m$^{3}$, the surface tension of sapphire
$\gamma$ is 0.91 J/m$^{2}$ (Ref. 28), and the ion flux is
2$\times10^{18}$ ions/m$^{2}$/sec. The ion-enhanced surface
viscosity $\eta_{s}$ is estimated to be 7.5$\times10^9$ Pa-s at an
ion energy of 600 eV.  This value is chosen in order to give quantitative agreement between the model and the experimental values of $\ell_x$.

The fitted power law coefficient of p=0.71 in Fig. 6(a) and that of
the IVF model (p=0.67) for the dependence of the wavelength on ion
energy at low temperature are indistinguishable within experimental
error. Moreover, we have observed that the RHEED pattern for the
sapphire (0001) surface disappears upon ion irradiation at room
temperature, confirming that ion bombardment
amorphizes the surface, or  at least induce a layer with a very high
defect density. Similar behavior of surface amorphization under ion
bombardment has been noted in the study of ion sputtered Si and
InP.~\cite{Frost41, Hazra43} The key idea of the IVF model is that
this thin layer can relax by a collective motion (``flow''), driven
by surface tension.

The SD model can be useful in predicting the high temperature ripple
wavelength.  The formula for the wavelength with its wavevector
along the x-axis can be expressed from Eq. (2), (4), (7) and (8),
as:
\\
\begin{equation}\label{wavelength ion energy dependence3}
  \ell_{x}\sim\sqrt{\frac{ncK_{SD}}{JdY(\theta)}}~,
\end{equation}
\\
where only d and Y($\theta$) are dependent on ion energy. Thus, the
SD model gives $\ell\sim\bm{\varepsilon}^{p}$ where p=-0.55 for
sapphire ripples at high temperature. The SD model is consistent
with the energy dependence at high temperature in Fig. 6(b).  A more refined model that combines both the IVF and SD mechanisms is given in section IV-D, which is also compatible with the data in fig. 6(b). Other variations of the SD model that include ion-bombardment effects are discussed in section V.
We also note that the sapphire surface exhibits a well developed RHEED pattern after etching at 1000 K, indicating a higher degree of surface crystallinity at this temperature.

\subsection{Dependence of Ripple Wavelength on Ion Incidence Angle (Low Temperature) }

The observed wavelength-angle phase diagram for sapphire ripples
produced by 600 eV Ar$^{+}$ bombardment at room temperature is
displayed in Fig. 7. The wavelength of sapphire
ripples is varied through a remarkably   wide range ( 30 nm to 2
$\mu$m) by changing the incidence angle. Below 40$^{\circ}$, the
ripple wavelength is particularly sensitive to the incidence
angle, while the wavelength is relatively constant in the
middle range from 40$^{\circ}$ to 65$^{\circ}$. Ion incidence at an
angle larger than 70$^{\circ}$ rotates the orientation
of the ripples by 90$^{\circ}$.

The theoretical wavelength-angle phase diagram for sapphire ripples
produced by 600 eV Ar$^{+}$ bombardment at room temperature is also
shown in Fig. 7. Several variables in the expressions of Eqs.
(4)-(11) were calculated using SRIM simulations, including the  ion
penetration depth $d$,  the sputter yield Y($\theta$), and the
widths of the deposited energy distribution, $\sigma$ and $\mu$.
Here, the sapphire surface binding energy U$_{0}$ is taken to be 2
eV/atom, and the ion flux is 2$\times10^{18}$ ions/m$^{2}$/sec. We
estimate the ion-enhanced surface viscosity to be
$\eta_{s}=7.5\times10^{9}$ Pa-s, as discussed in Section B. The
predicted wavelengths $\ell_{x}$ and $\ell_{y}$, based on the SES
(the dash-dot line and dotted line) and IVF (the solid line and dashed
line) models, are compared with those experimentally
obtained.

There are three distinct regions in Fig. 7. Region I:
$\nu_{x}<\nu_{y}<0$, K$_{xx,SES}>0$, K$_{yy,SES}>0$, K$_{IVF}>0$.
This region spans from normal incidence to
oblique incidence around 62$^{\circ}$. The ripple
with wavevector parallel to the projection of the ion beam ($\ell_x$) dominates the
surface morphology in this region. For the SES model, the $\ell_{y}$ oriented ripple is predicted to dominate the morphology over most of region I, except near the region I/II boundary where the ripple orientation rotates by 90$^\circ$.  This is in clear disagreement with the experimental observations.  On the other hand, the IVF model prediction agrees with the observed ripple orientation over the whole range.  However, the IVF mechanism does not  predict the observed large wavelength-angle dependence below 40$^{\circ}$. This point will be discussed further in section V.

Region II: This region is characterized by negative K$_{xx,SES}<0$,
which prevents the appearance of $\ell_{x}$ for the SES model, and thus the
SES model predicts only  $\ell_{y}$ ripples in this region. In the IVF model, the $\ell_{x}$ wavelength increases to infinity at
the region II/III boundary near 65$^{\circ}$. Thus, the IVF model  predicts that the dominant ripples
will switch their wavevector orientation to the y-direction  in this
region. This boundary could be adjusted towards higher angles since it is very sensitive to the
change of simulated parameters, such as $d$,  $d_{\sigma}$ and
$d_{\mu}$.  The experimental observation is that $\ell_{x}$  ripples are still observed, but longer scale order in the orthogonal direction begins to build up.  Overall, the behavior in this region agrees reasonably well  with the prediction of the IVF model.

Region III: $\nu_{x}>0$, $\nu_{y}<0$, K$_{xx,SES}<0$,
K$_{yy,SES}>0$, K$_{IVF}>0$. The  $\ell_{x}$ ripple is not stable in either
model, since $\nu_x$ becomes an effective smoothing term when it is positive. Near 90$^{\circ}$, $\ell_{y}$ either drops to zero (SES) or increases to infinity (IVF).  Again, the IVF model correctly predicts the observed behavior, at least qualitatively.

Ex-situ AFM images in Fig. 8 display surface morphologies obtained
at different angles of incidence for ion sputtered sapphire,
corresponding to the observed phase diagram in Fig. 7. Figs. 8(a)
and 8(b) show images for off-normal incidence at 25$^{\circ}$, which
produces micron-scale ripples with wavevector parallel to the ion beam
direction, and are readily visible in the large-scale image [Fig.
8(b)]. In contrast, 55$^{\circ}$ incidence, shown in Fig. 8(c),
produces a well-ordered nanorippled surface with the wavevector
parallel to the projection of the incoming ion beams along the
surface, which has a wavelength similar to that shown in Fig. 3. We note
that at the larger scale in Fig. 8(d) the surface roughness is also
correlated with wavevector perpendicular to the incoming ion
beam, as predicted by $\ell_{y}$ in the phase diagram. Ion
incidence at 65$^{\circ}$ still creates detectable ripples
with wavevector parallel  to the projection of the incoming ion beams in Fig.
8(e), but obvious submicron furrows oriented along the ion beam
direction are observed  in Fig. 8(f). Grazing ion
incidence at 75$^{\circ}$ switches the orientation of the ripple
wavevector perpendicularly.  Fully developed ripples are observed, with an unusual rod-like
structure, as shown  in Figs. 8(g) and 8(h).

\subsection{Temperature Dependence of Ripple Wavelength}

Fig. 9 shows the observed ripple wavelength dependence on inverse
temperature 1/T for two different angles of incidence. The ion
energy is 600 eV for both angles of incidence. All samples
are preheated to a chosen temperature and then sputtered at this
temperature until a well-defined wavelength is established. The
ripple wavelength obtained at 45$^{\circ}$ is constant at low
temperature and increases significantly when the temperature increase
over 700 K.

We have used the SD mechanism to  describes the temperature
dependence of the ripple wavelength at 45$^{\circ}$. The wavelength
varies as $\ell\sim(T)^{-1/2}exp(-\Delta{E}/2k_{B}T)$, where
$\Delta{E}$ is the activation energy for surface diffusion and $k_B$
is Boltzmann's constant.~\cite{BH10} However, this expression 
does not take into account  the low-temperature component of ion bombardment induced smoothing. Thus in Fig. 9, the observed
dependence of the ripple wavelength on temperature at the incidence
angle of 45$^{\circ}$ is modeled  with a modified $K_{xx}$ (solid line), as:

\begin{equation}\label{wavelength temperature dependence1}
     K_{xx}(\varepsilon, \theta, T)=K_{IVF}(\varepsilon,
\theta)+K_{SD}(T)~.
\end{equation}

Taking Eq. (15) and the expression for K$_{SD}$ in Eq. (8), we can
obtain the temperature dependence of the ripple wavelength from Eq.
(2). The fitting parameters are the activation energy $\Delta{E}$
for surface diffusion and the areal density of mobile species $\rho$
involving in surface diffusion. An activation energy of $\Delta{E} =
0.72$ eV is extracted from Fig. 9 (solid line). Taking
n=1.17$\times10^{29}$ atoms/m$^{3}$, $\gamma$=0.91 J/m$^{2}$, and
assuming D$_{0}$ with the order of magnitude 10$^{-8}$ m$^{2}/s$, we
can estimate $\rho$ to be of the order of 10$^{13}$
m$^{-2}$.

The calculated wavelengths for 35$^{o}$ incidence based on Eq. (2), (4), (8), (11) and (15) are shown by the dashed line in Fig. 9.  A weaker, but still significant temperature dependence is predicted, which is not observed experimentally (square symbols). Rather, the experimental ripple wavelengths at 35$^{\circ}$ are independent of temperature.  This indicates that the different smoothing mechanisms (i.e. thermal vs. non-thermal mechanisms) have different dependences on the angle of incidence. In particular, the rapid increase in wavelength at low angles  is inferred to be due to a non-thermal smoothing mechanism that increases at low angles to dominate over the other mechanisms, but is not included in our present model.

Finally, Fig. 10 shows the observed wavelength-angle phase diagram
($\ell_{x}$ vs. $\theta$) for sapphire ripples obtained at two
temperatures, 300 K and at 1000 K. The theoretical curves based on
Eq. (2), (4), (8), (11) and (15) are also plotted. The solid line for 300 K is
identical to the solid line in Fig. 7, which is just based on the
IVF model. The dashed line for 1000 K includes both SD and IVF
smoothing effects, combined as shown in Eq. (15). We note that the
experimental ripple wavelength $\ell_{x}$ exhibits a very large increase
and is not sensitive to thermal activation for incidence angles below
40$^{\circ}$. Taken together, these observations again indicate a
strong  non-thermal smoothing mechanism which is not
adequately explained by any of the models under consideration.

\section{\label{secLevel1}Discussion}

The observations in the section IV-B and IV-C indicate that the IVF and SD
models fits some of the trends of the wavelength dependence on
experimentally accessible parameters.
However, some of the characteristics during the ripple formation are
beyond the current theories as described in sections IV-C and IV-D.
Fig. 7 shows that the obtained wavelength  for
incidence angle lower than 40$^{\circ}$ spans a range of two orders
of magnitude from 30 nm to 2 $\mu$m. Figs. 8(c) and 8(d) confirm
that ion incidence at  25$^{\circ}$ only produces surface features  at the larger scale. Furthermore, ion sputtering at normal
incidence does not roughen the surface at all from the onset of
irradiation, which is confirmed by a real-time GISAXS study. This is
in contrast to the theoretical behavior which predicts that the
ripple wavelength only increases slightly as normal incidence is
approached. If the nonlinear terms $\lambda_{x}$ and $\lambda_{y}$
are introduced into the continuum equation describing the surface
 motion, kinetic roughening~\cite{Makeev2} is expected in
the region I of the phase diagram (when $\lambda_{x}\cdot
\lambda_{y}>0$), but such roughening is not observed. These unusual effects lead
us to propose that there is an additional smoothing mechanism that
dominates the behavior near normal incidence.

We have considered the fact that the theoretical ripple wavelength
is very sensitive to the ion range $d$, so that a small (factor of
two) uncertainty in $d$ would have a large (order of magnitude)
effect on the calculated wavelength for certain models. A factor
that is not taken into account in SRIM simulations is the
ion-channeling effect.~\cite{Feldman37} Ion channeling in
crystalline surface layers would be strongest near low-index
crystallographic directions, and could potentially lead to an
increase in ion range $d$ near the [0001] axis, and a corresponding increase in ripple wavelength. Based on RHEED
analysis of sapphire surfaces after ion bombardment at different
temperatures,  there is a striking increase in
surface order at higher substrate temperatures, and we would expect a stronger
channeling effect for the experiments at high
temperatures than at low temperatures. However, Fig. 10 does not
show any effect of temperature at the lowest angles of incidence,
effectively refuting this idea.  Furthermore, the cutoff of
ion-channeling by a low energy threshold in the range between 0.1-1
keV may weaken its possible influence on the formation of erosion patterns.~\cite{Hobler53}

Another important factor that is not taken into consideration by
current models to explain surface morphology created by ion
sputtering is ion impact induced lateral mass
redistribution.~\cite{Carter29,Johnson54,Moseler51} It can induce a
form of surface smoothing process that is different from the SD,
SES or IVF relaxation mechanisms previously discussed. Impact-induced downhill
currents have been identified as the driving force underlying the
ultra-smoothness of surfaces resulting from ion assisted film deposition.~\cite{Moseler51}   The
ion induced lateral currents gives rise to curvature dependent ($\partial^2h/\partial x^2, \partial^2h/\partial y^2$)
smoothing terms, which can weaken or even cancel the curvature
dependent roughening term in Eq. (1) [see Eq. (6) in Ref. 4]. 
No ripples can form under conditions where this type of smoothing
term is dominant, since the wavelength dependence  exactly matches that of the prevailing roughening mechanism. Thus, the idea that this additional smoothing
mechanism plays a role at low angles of incidence provides a route
which may explain the anomalous wavelength-angle phase
diagram in Figs. 7 and 10.  We also note that the
lateral current term is expected to be strongest at low angles of
incidence.~\cite{Carter29}

The behavior of the ripple formation at high temperature can
be explained reasonably by the SD model. However, another important
fact observed in our experiments, but not discussed above, is that thermal annealing at 1000 K
without ion irradiation does not produce any distinguishable decay
of the amplitude of as-prepared ripples, in contrast with previous
studies on Si and Ag surfaces.~\cite{Erlebacher46, Pedemonte47} This
indicates that the surface smoothing at high
temperature is not by a type of surface diffusion that is purely
thermally activated. Rather, it is likely to involve to the creation of mobile species on the surface during ion
bombardment. Further work on the flux
dependence of the $\ell_{x}$ at high temperature will assist us in
clarifying the dominant creation process for mobile species
underlying the ion-enhanced surface diffusion.~\cite{Erlebacher4}

\section{Summary}
In summary, the formation and characteristics of ripple morphologies
on sapphire surfaces produced by ion sputtering are systematically investigated
 by  \textsl{in-situ} GISAXS and \textsl{ex-situ} AFM. The ripple
wavelength can be modulated effectively in a wide range of 20 to
2000 nm by changing the ion incidence angle, ion energy and
temperature. This phenomenon provides an easy route to fabricate 
nanostructured surfaces for exploring  novel 
nanoscale phenomenon.  The IVF and SD smoothing mechanisms are shown to play an important role in the formation of the sapphire ripple structure.
 The possible importance of impact-induced lateral currents as a smoothing mechanism should also be
investigated further.
\\
\\
                      \textbf{ACKNOWLEDGMENTS}
\\

We wish to acknowledge the experimental assistance and facility
supports of Dr. Lin Yang from NSLS-X21  and Prof. Jie Yang from the
UVM physics department. This material is based upon work supported by the
National Science Foundation under Grant No. 0348354 and by the
Department of Energy under Grant No. DEFG0203ER46032.

\newpage
\textbf{Figure Captions}

FIG 1. Schematic of the x-ray scattering geometry. The
z-axis is always taken to be normal to the sample surface, and  the incident x-ray beam defines the y-z plane.
$\textbf{k}_{\textbf{i}}$ and $\textbf{k}_{\textbf{f}}$ represent
the incident and scattered wave vectors, respectively. The
components (Q$_{x}$, Q$_{y}$ and Q$_{z}$) of the scattering momentum
transfer
$\textbf{Q}=\textbf{k}_{\textbf{f}}-\textbf{k}_{\textbf{i}}$ are
defined by the glancing angles $\alpha_{i}$, $\alpha_{f}$ and
in-plane (x-y plane) angle $\psi$. The linear PSD is oriented with its long direction in the plane of the surface. 

FIG 2. Grazing incidence small angle x-ray scattering (GISAXS)
patterns acquired from Q$_{x}$-Q$_{z}$  scans. Asymmetric GISAXS
patterns indicate the shape anisotropy on the ripples. The bar of
gray shades represents the logarithmically scaled intensity.

FIG 3. A 500 nm by 500 nm  AFM image taken
after the GISAXS measurements.  The ripple wave vector is parallel to the projected ion beam
direction.  In the text, this is referred to as the $\ell_x$ orientation of ripples.

FIG 4. Time-resolved GISAXS measurements indicate the increase of the lateral correlations on a sapphire surface during ion exposure. The surface ripples are produced by 500 eV Ar$^{+}$
bombardment at 45$^{\circ}$  incidence. The vertical
component of the scattering momentum transfer Q$_{z}$ was fixed at
0.92 nm$^{-1}$ and Q$_y$ was offset to $3 \times 10^{-3}$ nm$^{-1}$.

FIG 5. An asymmetric saw-tooth model profile ($\alpha<\beta$). The
parallel component of the incident ion beam along the substrate
surface is along the -x direction, as defined in Fig. 1. The term
$\partial{h}/\partial{x}$ represents the local slope of ion etched
surface curvature. The magnitude of $\alpha$ and $\beta$ are exaggerated for clarity.

FIG 6. The dependence of the ripple wavelength ($\ell_{x}$) of ion
etched sapphire on the incident ion energy ($\varepsilon$). (a) 300 K substrate temperature. The square symbols are for the high flux ion source and the circle symbols are for the low flux ion source. The solid line is the curve for the best fit exponent of p=0.71. For comparison, curves for p=1 and p=0.5 are plotted as dashed and dotted
lines, respectively. (b) 1000 K substrate temperature. The solid line is for the
best fit exponent of p=-0.44.  For comparison curves for  p=-0.25 and p=-0.75 are plotted as dashed and
dotted lines, respectively.

FIG 7. The wavelength-angle phase diagram for sapphire ripples
produced by 600 eV Ar$^{+}$ bombardment. Circle symbols: experimental
wavelength $\ell_{x}$ at room temperature. Square symbol: experimental
wavelength $\ell_{y}$ at room temperature (Only one point at
75$^{\circ}$ is shown). The error bars are comparable to the height
of the symbols. Two sets of curves are shown:  The upper set is for the IVF model, with the solid line representing the simulated $\ell_{x}$ wavelength, and the dashed line representing the $\ell_{y}$ wavelength.  The lower set of curves is for the SES model, where the dot-dashed line represents the  $\ell_{x}$ wavelength and the dotted line represents the $\ell_{y}$ wavelength.
Experimentally, the three regions of the phase diagram represent the ranges of angles where $\ell_{x}$  and $\ell_{y}$ are dominant (I and III, respectively), and a narrow transition region ( region II). The SES model predicts that $\ell_y$ ripples dominate in all three regions, which is not consistent with the data.  The IVF model gives better agreement with the data, as described in the text.

FIG 8.  AFM images of surface morphology at
different angles of incidence: (a) and (b) are for an incidence
angle of 25$^{\circ}$, (c) and (d) are for an incidence angle of
55$^{\circ}$, (e) and (f) are for an incidence angle of
65$^{\circ}$, and (g) and (h) are for an incidence angle of
75$^{\circ}$. Two different image sizes are shown for each angle of
incidence. The white arrow indicates the projection of the ion beam direction along the surface.  The progression of ripple orientation and wavelength can be seen in this sequence of images.  In particular, we note the large scale ripples at 25$^\circ$ in (b), followed by much smaller wavelength ripples at 55$^\circ$ in (c) and 65$^\circ$  in (e), and finally the rotation of the ripple orientation in (g) and (h).

FIG 9. The wavelength $\ell_{x}$ as a function of inverse
temperature at different angles of incidence and constant ion energy of
600 eV.  Theoretical curves based on Eqs. (2), (4), (8), (11) and (15) are shown for the set
of ripples wavelength $\ell_{x}$ obtained at incidence angles of 35$^{\circ}$ (dash
line) and 45$^{\circ}$ (solid line) off-normal incidence.

FIG 10. The wavelength-angle phase diagram for sapphire ripples
produced by 600 eV Ar$^{+}$ bombardment at different temperatures.
The error bars are comparable to the height of the symbols. The
curves for  $\ell_{x}$ based on Eqs. (2), (4), (8), (11) and (15) are compared with
the measured ripples wavelengths obtained at 300 K (solid line) and
1000 K (dash line), respectively.
\end{document}